# A High-Quality Thermoelectric Material Database with Self-Consistent ZT Filtering

## Authors

Byungki Ryu,[1,2,*] Ji Hui Son,[1] Sungjin Park,[1,†] Jaywan Chung,[1] Hye-Jin Lim,[1,†] SuJi Park,[1,†] Yujeong Do,[1,†] and SuDong Park[1]

## Affiliation

1 Energy Conversion Research Center, Electrical Materials Research Division, Korea Electrotechnology Research Institute (KERI), Changwon, 51543, Republic of Korea

2 Electric Energy and Material Engineering, School of KERI, University of Science and Technology, Changwon, 51543, Republic of Korea

[*]Correspondence: byungkiryu@keri.re.kr

[†]These authors were affiliated with KERI at the time of this work, but are no longer at the institution.

# Abstract

This paper presents a curated **t**hermo**e**lectric **mat**erial **d**ata**b**ase, **teMatDb,** constructed by digitising literature-reported data. This database includes temperature-dependent thermoelectric properties (TEPs), such as Seebeck coefficient, electrical resistivity, thermal conductivity, and figure of merit (ZT), along with metadata on materials and their corresponding publications. A *self-consistent* ZT (**Sc-ZT**) filter set was developed to measure ZT errors by comparing the reported ZT values from figures with ZT values recalculated from the digitised TEPs. Using this Sc-ZT protocol, we generated **tMatDb272**, comprising 14,717 temperature-property pairs from 272 high-quality TEP sets across 262 publications. This method identifies various types of ZT errors, such as resolution errors, publication bias, ZT overestimation, interpolation and extrapolation errors, and digitisation noise, and excludes inconsistent samples from the dataset. **teMatDb272** and the **Sc-ZT** filtering framework offer a robust dataset for data-driven and machine-learning-based materials design, device modelling, and future thermoelectric research.

**Version log**

(2025-May-25) Version 1 uploaded.

(2025-May-27) Version 2 uploaded.

(2025-July-25) In version 3, typo was corrected for the last line in Table 1

## 1. Background and Summary

High-quality material data accelerate scientific discovery, enable predictive modelling, and support experimental design, particularly in emerging fields such as materials informatics and machine-learning-driven materials science[1–6]. For example, calculation datasets such as the Materials Project[2] and Omat24[7] have paved the way for the development of foundational machine-learning interatomic potentials[8–10]. Moreover, the 3D structure spaces of proteins have been predicted and explored using AlphaFold, extending structural coverage far beyond experimental efforts[11].

**However, experimentally measured data for many functional materials remain scarce compared to calculation data**[4]**.** For example, the recently released OMat24 dataset includes approximately 118 million inorganic material structures labelled with energy, force, and strain data[7], whereas the world's largest experimentally determined inorganic crystal structure database, ICSD, contains only 318,901 structures as of March 31, 2025[12]. Furthermore, existing databases (DBs) often do not prioritise functionality and contain errors and inconsistencies[13]. These issues hinder the reliability of data-driven research and limit the effectiveness of machine learning models, whose predictive power depends critically on the quality of the observed data and the representativeness of the explored data space[14,15]. This poses a major obstacle to future research aiming to integrate domain knowledge and data science[4,14].

**Many functional materials face considerable challenges related to data size,**

**inconsistency between reports, and quality. Among the most impactful efforts to address the experimental data size challenges is Starrydata2**[6], the world's largest open thermoelectric DB; as of May 1, 2025, it contains over 53,000 samples from more than 9,000 publications. This DB has greatly improved accessibility to experimental thermoelectric material data and has even evolved into a general-purpose DB for functional materials such as magnetic, battery, and thermal insulating/conducting materials. However, Starrydata2 includes digitisation noise, inconsistencies, and outliers, mainly due to unit and labelling errors. In our previous study on best-efficiency exploration, we applied 33 distinct physics filters, including noise, size, physics, and Lorenz filters, to clean the Starrydata2 dataset[13]. For example, in the 20211124 version of the dataset, 3,121 of 16,420 valid samples were filtered out through this process, which we refer to as **Classical filtering**. These efforts do not reflect a flaw in the DB itself, but rather highlight the need for standard thermoelectric data filtering protocols and quality assessment methods as community databases grow in scale and scope. Therefore, we aim to complement these valuable open resources by constructing a curated, high-quality, and self-consistent thermoelectric materials database to further support reliable and data-driven future research.

**To address these challenges, we present teMatDb, a curated thermoelectric materials database constructed from literature-reported temperature (T)-dependent thermoelectric property (TEP) data.** Each sample dataset includes the Seebeck coefficient ($\alpha$), electrical resistivity ($\rho$), thermal conductivity ($\kappa$), and the figure of merit (ZT), all digitised from published figures. A key feature of teMatDb is the implementation of a **self-consistent ZT (Sc-ZT) filtering** method, in which reported ZT values from figures ($\mathrm{ZT}_{\mathrm{fig}}$) in publications are compared against recalculated ZT values derived from the digitised TEPs ($\mathrm{ZT}_{\mathrm{TEP}}$) using the relation $\mathrm{ZT} \coloneqq \alpha^2\rho^{-1}\kappa^{-1}T^1$. The ZT error, defined as $\delta(\mathrm{ZT}) \coloneqq \mathrm{ZT}_{\mathrm{fig}} - \mathrm{ZT}_{\mathrm{TEP}}$, can serve as

an error metric for detecting various types of digital and publication errors. By applying this protocol, we systematically identified and excluded erroneous data entries, resulting in a high-quality dataset named *teMatDb272*, which exhibits improved internal consistency between TEPs and ZT values.

**The three key thermoelectric properties ($\alpha$, $\rho$, and $\kappa$) play a fundamental role in determining the thermoelectric effects and energy conversion.** Thermoelectric devices directly convert thermal energy into electricity and vice versa via the Seebeck and Peltier effects that occur in materials subjected to a temperature gradient. In such materials, the electrical current density ($J$) and thermal current density ($J^Q$) follow the steady-state thermoelectric differential equations:

$$J = \sigma(E - \alpha \nabla T)\,, \tag{1}$$

$$J^Q = -\kappa \nabla T + \alpha T J\,. \tag{2}$$

where $\sigma$ is electrical conductivity ($\sigma \coloneqq \rho^{-1}$). TEPs are not constant but are T-dependent curves, which are essential not only for predicting device performance but also for extracting physical properties. For example, simple fitting of experimental data with a parabolic band model can yield the effective mass[16], density-of-states[17], band gap[18], Lorenz number[19], and electron relaxation time[20]. Therefore, accurate knowledge of T-TEP data pairs is highly valuable for understanding material systems and guiding future thermoelectric materials discovery and analysis.

The thermoelectric efficiency η of a leg can be described by:

$$\eta(T_c, T_h, J) = \frac{J\left(\int_c^h \alpha \, dT - J \int_h^c \rho \, dx\right)}{-\kappa_h \nabla T_h + \alpha_h T_h J}\,. \tag{3}$$

In practice, the ZT value, derived from the constant property model, has long served as a milestone in thermoelectric research, guiding material discovery and device design. Because ZT is calculated as a simple product of three TEPs and temperature, the reliability of published ZT directly depends on the internal consistency with its constituent properties. Ensuring this consistency is critical not only for assessing data quality but also for ensuring data transparency, especially in a research landscape where peak ZT values often carry disproportionate scientific weight. Although the temperature-dependent ZT curve itself is not a direct measure of device efficiency, it can serve as a valuable reference. Compared with digitised TEPs, ZT enables cross-validation of data integrity and helps identify potential digitisation errors. Thus, ZT serves as a reference for evaluating data consistency, integrity, and fidelity.

**The final curated dataset, teMatDb272, contains 14,717 temperature-property pairs from 272 samples high-quality TEP sets across 262 publications.** In detail, the dataset includes 3,853 entries (rows) each for Seebeck coefficient $\alpha(T)$ and electrical resistivity $\rho(T)$, 3,422 for thermal conductivity $\kappa(T)$, and 3,589 for ZT. **Figure 1** illustrates the distribution of these curated thermoelectric data with the temperature and thermoelectric materials space. Although the dataset size is smaller than that of starryz10840, which is a newly curated dataset from Starrydata2 (version **starrydata_dataset_250501-0300** as of May 2025) using the Sc-ZT filtering protocol, teMatDb272 shows comparatively excellent data coverage. The included TEP data are well distributed over a wide range of ZT values while preserving representativeness across the main thermoelectric property data space. Notably, **teMatDb272** contains a few high-ZT data points that are not currently included in starryz10840. This suggests that **teMatDb272** may complement **Starrydata2** for high-performance material exploration.

## 2. Methods

### TEP data digitisation

Thermoelectric properties are temperature-dependent curves and are typically measured over a broad temperature range. However, in most experiments, TEP measurements are performed only at a limited number of discrete temperature points. As a result, the reported T-TEP data pairs are often presented as figures in publications rather than as tables. To digitise TEP data pairs, we used two tools: WebPlot Digitizer (https://automeris.io/)[21] and Plot Digitizer (https://plotdigitizer.sourceforge.net/)[22].

**Figure 2(a-c)** show examples of TEP data digitised from figures for one of the material samples in teMatDb272 (sample_id = 43)[23]. The digitised data points were connected using piecewise linear interpolation to generate continuous TEP curves. Although polynomial fitting could be applied, it might distort the original values at the measured temperature points. To preserve the originality of the measured values, we employed interpolation instead of regression-based fitting.

Every digitisation process may introduce errors. Because ZT can be directly calculated from three TEPs ($\alpha$, $\rho$, and $\kappa$) it is possible to compare the $\mathrm{ZT}_{\mathrm{fig}}$ values with the $\mathrm{ZT}_{\mathrm{TEP}}$ values, as shown in **Figure 2(d)**. To evaluate the consistency between $\mathrm{ZT}_{\mathrm{fig}}$ and $\mathrm{ZT}_{\mathrm{TEP}}$, we define and calculate the ZT error $\delta(\mathrm{ZT}) \coloneqq \mathrm{ZT}_{\mathrm{fig}} - \mathrm{ZT}_{\mathrm{TEP}}$ over the measured temperature range. However, the digitisation temperature points of the ZT and TEP data do not always coincide. To overcome this, we generated collocated TEP datasets at intervals of 2 K using piecewise linear interpolation of $\alpha(\mathrm{T_i})$, $\rho(\mathrm{T_j})$, $\kappa(\mathrm{T_k})$, and $ZT(\mathrm{T_l})$. Note that we interpolated electrical resistivity rather than electrical conductivity, considering how resistivity directly contributes to the device's ZT expression over a wide temperature range[24–27].

**Figure 2(e)** shows the ZT error for sample_id = 43. The absolute ZT error is less than 0.03, whereas the peak ZT value is 1.2. Thus, the relative ZT error compared to the peak ZT was less than 2.5%. **Figure 2(f)** shows the quantile-quantile (Q-Q) plot of $\delta(\mathrm{ZT})$ to evaluate its distribution. The theoretical quantile was calculated from a normal distribution, and the y-axis represents the ZT errors sorted by magnitude. Here, the $R^2$ value is 0.9439, indicating that the error distribution closely follows a normal distribution, with a slight positive bias of approximately 0.02. This bias suggests that the ZT values reported in the figure tend to be slightly larger than those recalculated from the digitised TEPs.

## Digitisation noise

To evaluate digitisation-induced noise, we generated a test figure (**Figure S1**) containing a set of randomly distributed points within a figure size of 3.0 × 3.5 inches. We then digitised the data and compared the digitised values with the original data. The digitisation results were found to be highly accurate relative to the resolution of the figure. The relative mean and relative maximum errors, calculated with respect to the peak value, were less than 0.5% and 1%, respectively, and the standard deviation was below 0.3%. These results indicate that, in general, digitisation can be performed with high fidelity under typical figure resolutions.

However, in the ZT reconstruction, both the electrical resistivity and thermal conductivity are used in reciprocal form. When the digitised value is close to zero, the reciprocal operation can significantly amplify the error. To evaluate this effect, we compared the reciprocal of the digitised values with that of the original data. We observed that when the original value was near zero, the relative maximum error could reach up to 8%. The details of this analysis are presented in **Supporting Tables S1 and S2**. These findings highlight the importance of resolution and value magnitude in determining digitisation accuracy, especially when the data are small and used in reciprocal form during ZT calculations.

**Types of ZT errors in digitised datasets**

For some material samples, the ZT error was negligible. However, in other cases, the error was substantial and could not be ignored. In this subsection, we categorise ZT errors into several types based on their observed characteristics: **resolution error**, **publication bias error, ZT overestimation induced by fitting, extrapolation error, interpolation error, and digitisation noise**.

The first error type is **resolution error**. One example is sample_id = 28[28], where the electrical conductivity is presented on a logarithmic scale and multiple data points/curves visually overlap. In this special case, we digitised more than three times to improve the accuracy. As a result, the ZT error $\delta(\mathrm{ZT})$ was reduced to below 0.08 (**Supporting Figure S2**).

The second error type is **publication bias error**, where the average ZT error over a wide temperature range is nonnegligible between $\mathrm{ZT}_{\mathrm{fig}}$ and $\mathrm{ZT}_{\mathrm{TEP}}$. **Supporting Figure S3** shows the digitised TEP data for sample_id = 95[29]. In this case, $\mathrm{ZT}_{\mathrm{fig}}$ is consistently and significantly larger than $\mathrm{ZT}_{\mathrm{TEP}}$ across the temperature range, with a $\delta(\mathrm{ZT})$ value of approximately 0.2, ten times higher than that of sample_id = 43 (**Figure 2**). We refer to this as a ZT bias error because the deviation in ZT values is consistently and significantly biased. Notably, both samples were reported by our institute, yet the data quality differs: one shows good agreement with $\delta(\text{ZT}) \approx 0.02$ (**Figure 2**), while the other shows a significant discrepancy with $\delta(\text{ZT}) \approx 0.2$ (**Supporting Figure S3**). A similar bias is observed for sample_id = 415 (**Supporting Figure S4**)[30]. The recalculated peak ZT is approximately 2.0, but the ZT reported in the figure reaches 2.4. This type of bias error is critical not only because it reflects data inconsistency, but also because high ZT values are widely used as performance metrics in material selection. As researchers often search for and rank materials based on their reported ZT values, a positive bias may have a more pronounced impact than a negative bias, potentially leading to misinterpretation and misplaced research attention.

The next error type is **ZT overestimation error,** induced by curve fitting beyond the measured temperature ranges. For some materials, such as sample_id = 113[31], the digitised TEP data exhibited noticeable noise in the raw data (**Supporting Figure S5**). However, the origin of this noise is unclear. It may be due to material instability or measurement instability. In some publications, such data have been smoothed by polynomial or spline fitting, which can lead to inaccurate ZT values, particularly at the edges of the temperature range. Although Chebyshev node interpolation might preserve data originality[32], a high-degree polynomial over equidistance nodes can deviate significantly from the original points. For sample_id = 113, the ZT curve reported in the figure appears smooth, whereas the ZT calculated from the digitised TEPs shows strong fluctuations. As a result, $\delta(\text{ZT})$ reaches up to 0.2. It is important to note that in this evaluation, $\delta(\text{ZT})$ is evaluated only within the overlapping temperature range between the ZT and TEP data. Although the raw TEPs may include digitisation noise, over-smoothing through curve fitting may artificially inflate the peak ZT, leading to misleading conclusions if the fitted curve is trusted more than the original ZT data.

Another critical type of ZT error is **extrapolation error**, as illustrated in **Supporting**

**Figure S5**. For sample_id =113, the peak ZT value reported in the published figure is 1.5, whereas the ZT recalculated from the digitised TEPs is only 1.07. This discrepancy arises from extrapolation beyond the measured temperature range. The TEPs were measured between 303.66 and 851.99 K, whereas the reported ZT curve spans from 303 to 898.45 K. The large temperature mismatch indicates that the peak ZT in the publication is based on an extrapolated segment. Because peak ZT values often occur near the highest measured temperatures, such extrapolation can significantly overestimate the performance. In this case, the extrapolated region contributes directly to a non-self-consistent peak ZT value, highlighting the importance of clearly distinguishing between measured and extrapolated data in thermoelectric reporting.

**Interpolation error** is particularly likely to occur near phase transitions or phase transformations, where the TEP curves may exhibit discontinuities. Because most TEPs are measured above room temperature, materials may become unstable in certain temperature regions. Consequently, around the phase-transition temperature, data points are often sparse or entirely absent, making interpolation unreliable. One example is a $Cu_2Se$-based thermoelectric material (sample_id = 300, **Supporting Figure S6**). Although $\delta(\mathrm{ZT})$ is relatively small near the peak-ZT temperature, it varies significantly near the phase transition. This highlights the limitation of linear interpolation when applied to non-smooth regions of the TEP curve, especially in thermal phase transition or transformation material systems.

**The final error type is digitisation noise,** which is unavoidable because the data shown in the figures occupy a finite visual space. In some cases, as discussed in the previous subsection (**Digitisation Noise**), when a digitised TEP value is very close to zero, the reciprocal operations involved in the ZT calculation can amplify small errors, resulting in large ZT errors. Moreover, it can overlap with resolution error, making it difficult to determine the true origin of the discrepancy. However, this noise is fundamentally introduced by the digitisation process itself.

We also observed **negative ZT errors** in a few cases. For example, the digitised values of the temperature and ZT can appear negative when the original TEP value is near zero. The ZT value should be zero or positive, but may be misinterpreted as slightly negative owing to digitisation artefacts. In the case of temperature, such errors can also arise from unit confusion in the original article, for example, when a Celsius scale is mistakenly labelled or used instead of Kelvin. We excluded these unphysical cases from our dataset.

**Sc-ZT filter development.**

During the construction of **teMatDb**, we encountered numerous cases in which the ZT error was nonnegligible. To address this issue, we performed double digitisation of all sample data in teMatDb v1.1.6. For certain large ZT error cases, the data were digitised three times. In addition, to minimise potential human error, three thermoelectric specialists with PhD degrees and more than five years of experience participated in the second and third rounds of digitisation. To systematically handle these inconsistencies, we developed a self-consistent ZT filtering protocol. Based on $\delta(\mathrm{ZT})$, we defined and calculated six types of ZT errors, as summarised in **Table 1**.

In many cases, the ZT values were reported in the same temperature range as the TEP measurements. However, in some cases, we observed temperature range mismatches, resulting in significant bias errors such as publication bias and extrapolation errors. To overcome this, we implemented the average ZT (Avg ZT) and Peak ZT filters, which evaluate ZT over defined temperature intervals in the figure or in TEPs. To handle errors arising from interpolation and digitisation noise, we introduced the maximum ZT error ($\mathrm{Max}\big(\delta(\mathrm{ZT})\big)$) filter and the root-mean-squared ZT error (Rms$\left(\delta(\mathrm{ZT})\right)$ filter, both of which are calculated over the overlapping temperature range between the ZT curve from the figure and the digitised TEPs. Finally, to ensure robustness in low-ZT regions, we defined two normalised $\mathrm{Max}\left(\delta(\mathrm{ZT})\right)$ filters: one relative to the Avg ZT and the other relative to the Peak ZT. This may help capture inconsistencies, especially interpolation errors.

**Sc-ZT filtering protocol**

A multistep filtering protocol was implemented based on the six Sc-ZT filters developed in this study. First, the raw TEP data were digitised. Next, collocated TEP curves were generated at 2 K intervals. We then computed $\delta(\mathrm{Avg\ ZT})$ and $\delta(\mathrm{Peak\ ZT})$ using ZT or TEP temperature rannges. We also computed $\delta(\mathrm{ZT})$ as a function of temperature for each sample across the collocated temperature range. Finally, Sc-ZT filtering was applied using a defined set of threshold criteria. The default Sc-ZT filtering criteria were set to (0.1, 0.1, 0.1, 0.1, 0.2, 0.2), as listed in **Table 1**. Increasing the criterion values allows more data to pass through, but potentially at the cost of reduced ZT-TEP consistency. Conversely, smaller criterion values improve self-consistency, but reduce the number of passed samples.

**Figure 3(a) and (b)** show the ZT-ZT comparison plots before and after Sc-ZT filtering. Before filtering, 355 digitised samples are present (**Figure 3(a)**). The $\mathrm{ZT_{fig}}$ values deviate significantly from the $\mathrm{ZT_{TEP}}$ values. After filtering, 272 samples remain, and the ZT-ZT plot (**Figure 3(b)**) aligns more closely along the diagonal, indicating improved ZT consistency. **Figure 3(c) and (d)** show the quantile-quantile (Q-Q) plots of $\delta(\mathrm{ZT})$ before and after filtering. Before filtering, $\delta(\mathrm{ZT})$ exhibits strong non-normal behaviour, ranging from 0.4 to below -0.8. However, after filtering, the $\delta(\mathrm{ZT})$ distribution approximates a normal shape with a high $R^2$ value of 0.9324. This demonstrates that filtering effectively removes anomalies such as bias, extrapolation, and interpolation error, leaving primarily digitisation noise.

## 3. Data Records

### Key data files, structure, and description

We digitised the thermoelectric property curves, applied the Sc-ZT filtering protocol using the default criteria (0.1, 0.1, 0.1, 0.1, 0.2, 0.2), and generated the final curated dataset, **teMatDb272**. In addition to the TEP curved data, we also compiled material-level metadata describing the composition, dimensionality, and processing information. In total, teMatDb272 contains four files. **Table 2** summarises the file types, filenames, descriptors, and associated contents.

### Data statistics for teMatDb272

**Table 3** summarises the database version information and key statistics for **teMatDb272**, including the update timestamp, Sc-ZT filtering criteria, temperature grid for property interpolation, and number of entries for each thermoelectric property. The dataset was generated from the mother DB teMatDb v1.1.6. In line with our commitment to transparent and open data practices, we provide detailed information on dataset structure and statistical coverage.

### TEP distribution and visualisation

**Figure 4** illustrates the distribution of the thermoelectric properties and visualises the complex relationships among the TEPs. **Figure 4(a) and (b)** show the relationship between the Seebeck coefficients and electrical conductivity, where the colour scale represents the power factor (PF) and ZT. Optimally high ZT values tend to occur at relatively lower electrical conductivities and higher Seebeck coefficients compared to those for high PF values. **Figure 4(c)** shows the correlation between thermal conductivity and electrical conductivity. In contrast, **Figure 4(d)** shows a largely scattered relationship between thermal conductivity and the Seebeck coefficient. **Figures 4(e) and 4(f)** visualise ZT versus PF and thermal conductivity, respectively, with the colour indicating the temperature. These plots reveal that high ZT values are primarily observed at elevated temperatures. Although high PF or low thermal conductivity generally favour high ZT, the distributions appear decoupled, highlighting complex trade-offs and material-dependent trends. A deeper analysis of these correlations is beyond the scope of this study and will be addressed in future work.

## 4. Technical Validation

### Technical Validation of Sc-ZT filter: (1) case of teMatDb

**Table 4 and Supporting Figures S7-S10** summarise the relationship between Sc-ZT filtering criteria, the number of retained samples, and the resulting $R^2$ values in the Q-Q plots of $\delta(\mathrm{ZT})$. Before Sc-ZT filtering, the dataset contains 355 samples, with $R^2 = 0.6864$ in the $\delta(\mathrm{ZT})$ Q-Q plot. Applying the default Sc-ZT filtering criteria of (0.1, 0.1, 0.1, 0.1, 0.2, 0.2) resulted in a curated set of 272 samples and an improved $R^2$ of 0.9324. Stricter criteria of (0.05, 0.05, 0.05, 0.05, 0.1, 0.1) reduced the number of samples to 187 and further increased $R^2$ to 0.9799. When applying even tighter criteria of (0.02, 0.02, 0.02, 0.02, 0.04, 0.04), only 71 samples remained, but $R^2$ reached 0.9845. These results indicate a clear trade-off: although stricter filtering effectively eliminates erroneous samples and improves self-consistency, it also significantly reduces the dataset size.

**Technical Validation of Sc-ZT filter: (2) case of Starrydata2**

To further validate the robustness of the Sc-ZT filtering protocol, we applied it to the Starrydata2 dataset as of May 1, 2025 (version: starrydata_dataset_250501-0300)[6]. From the raw dataset, we selected only a subset of property keys relevant to the thermoelectric properties. Specifically, we chose 'Temperature' for the x-axis, and 'Seebeck coefficient' as well as 'Electrical resistivity' and 'Electrical conductivity' for the y-axis to reconstruct the electrical transport properties. For thermal transport, we included 'Thermal conductivity', and for performance metrics, we used 'Power Factor' and 'ZT'. Although Starrydata2 contains additional property keys, we restricted our selection to these core fields to ensure consistency and focus in the Sc-ZT filtering application.

First, we extracted the non-duplicated sample_id. The SID of the duplicated sample_id cases were excluded. Next, we converted the file 'starrydata_curves.csv' into raw TEP data files. Following the same Sc-ZT filtering protocol used for teMatDb, we generated a collocated TEP dataset from samples for which both the TEP curves and ZT values were complete and valid. We then calculated the ZT error and applied two types of filters:

(1) the **Classical filter,** which was developed in our previous study[13], and

(2) the **Sc-ZT filter,** which was newly developed in this work.

Before filtering, Starrydata2 contained numerous erroneous ZT values, particularly abnormally high values caused by unit label inconsistencies, such as using $\mathrm{mV\ K^{-1}}$ instead of $\mathrm{\mu V\ K^{-1}}$ or vice versa. Because the Seebeck coefficient is squared in the ZT formula, such unit mismatches can result in $\mathrm{ZT_{fig}}$ values that are either $10^6$ times larger or $10^{-6}$ times smaller than the $\mathrm{ZT_{TEP}}$ values (**Figure 5(a)**). After filtering, both the Classical and Sc-ZT filtering methods effectively removed non-self-consistent ZT entries (**Figure 5(b)**). Moreover, we observed that the Sc-ZT filter was more effective at identifying subtle inconsistencies. Initially, the dataset consisted of 15,532 samples from 3,293 publications. After Classical filtering, it was reduced to 15,053 samples from 3,444 publications. Many erroneous samples were cleaned compared to the raw data of 20211124, as reported before[13]. When the Sc-ZT filtering was applied, the dataset was further reduced to 10,840 samples from 2,721 publications. Detailed statistics are provided in **Supporting Table S3**.

## Limitations of Sc-ZT filtering compared to Classical filtering

The **Classical filtering** method employs relatively loose criteria, making it less effective at identifying subtle but critical errors. Therefore, it may be preferred in applications in which maintaining a larger dataset is prioritised over strict data quality. Moreover, Classical filtering is based on detecting known error types, and any data that are not explicitly labelled as erroneous are implicitly treated as valid. Additionally, no quantitative measure of internal consistency is provided.

In contrast, the Sc-ZT filtering protocol introduced in this paper offers a physically grounded and tuneable framework for evaluating the consistency between the reported ZT and digitised TEPs. Its filtering strength can be controlled using numerical thresholds, making it particularly powerful in scenarios requiring high data reliability.

For example, in the machine-learning-based prediction of thermoelectric property curves, Sc-ZT filtering is expected to outperform Classical filtering owing to its enhanced control over transport and ZT data consistency. Although teMatDb272 is small, its high quality makes it suitable for use in machine learning model development. Furthermore, the Sc-ZT filtering method is generalisable and can be applied to other thermoelectric datasets to assess and improve data integrity.

## Usage Notes

### teMatDb272

The teMatDb272 dataset can be directly accessed and downloaded from the following GitHub repository and its correspondong Zenodo DOI:

GitHub: https://github.com/byungkiryu/teMatDb/tree/main/teMatDb_publication

DOI: 10.5281/zenodo.15518035

### Pre- and post-processing scripts

All digitisation scripts and the full pre- and post-processing code used for generating

**teMatDb272** are also provided in the GitHub repository:

GitHub: https://github.com/byungkiryu/teMatDb

DOI: 10.5281/zenodo.15518035

**Starryz dataset**

In addition to teMatDb272, we provide the starryz dataset, which was derived from Starrydata2 and filtered using both the Classical and Sc-ZT filtering protocols. Three versions are available: starryz10840, starrryz15053, and starryz15532, corresponding to different filtering states. Detailed statistics are provided in **Supporting Table S4**.

**Private link for starryz dataset:** https://figshare.com/s/50a78a58d6a84a5b6302 (will be updated with a public link after manuscript publication)

**teMatDb and sample ZT error visualisation**

An interactive web interface for visualising sample-level ZT-TEP consistency and $\delta(\mathrm{ZT})$ error is available at https://tematdb.streamlit.app/.

## Code Availability

All source code used for digitisation, filtering, and data processing is openly available on GitHub: https://github.com/byungkiryu/teMatDb. Each sample in teMatDb272 can also be interactively viewed via the Streamlit web interface: https://tematdb.streamlit.app/. Alternatively, the repository can be cloned using Git and run locally. All related figures and additional datasets, including those derived from Starrydata2 and used in the starryz datasets, are available at Figshare.com (public DOI or link will be assigned upon publication; currently, a private link is given as https://figshare.com/s/50a78a58d6a84a5b6302). For further assistance or access to specific files, please contact the corresponding author.

## Author Contributions

**Byungki Ryu:** Conceptualization, Methodology, Data curation, Software, Validation, Formal analysis, Supervision, Visualization, Writing – original draft, Writing – review & editing, Funding acquisition, Project administration. **Sungjin Park:** Data curation, Investigation, Validation. **Jaywan Chung:** Resources, Software, Data curation, Validation, Funding acquisition. **JiHui Son:** Investigation, Data curation. **Hye-Jin Lim:** Data curation. **SuJi Park:** Data curation. **YuJeong Do:** Software. **SuDong Park:** Project administration, Supervision, Funding acquisition.

## Competing Interests

The authors declare no competing interests.

## Acknowledgements

This work was supported by the Energy Efficiency & Resource Core Technology Program of the Korea Institute of Energy Technology Evaluation and Planning (KETEP) granted from the Ministry of Trade, Industry & Energy (MOTIE), Republic of Korea (Grant No. 2021202080023D). This work was also supported by the Korea Electrotechnology Research Institute (KERI) Primary Research Program through the National Research Council of Science and Technology (NST) funded by the Ministry of Science and ICT (MSIT) of the Republic of Korea (No. 25A01013). It was also supported by the National Research Foundation of Korea (NRF) grant funded by the Korea Government (MSIP) (2022M3C1C8093916). All sentences were originally written by the authors and later polished using ChatGPT (OpenAI) for grammar correction and improved clarity. All scientific content, analyses, and conclusions were solely developed and verified by the authors. We would like to thank Editage (www.editage.co.kr) for English language editing. S. Park, H.-J. Lim, S. J. Park, and Y. Do contributed to the early stages of data curation and digitization. They were affiliated with KERI during the project period.

**Table 1. Sc-ZT filters and description.** Summary of the Sc-ZT filtering scheme used in teMatDb272, including the ZT filter name, ZT error equation, target error to be filtered, and filtering criteria applied for error removal.

| ZT filter name | ZT error equation | Target error | Filtering criteria |
|---|---|---|---|
| Physics filter | $\mathrm{T} > 0$<br>$\mathrm{Avg}(\mathrm{ZT}) > 0,\ \mathrm{Peak}(\mathrm{ZT}) > 0$ | Unphysical | $< 0$ |
| Avg ZT filter | $\mathrm{Avg}(\mathrm{ZT}_{\mathrm{fig}}) - \mathrm{Avg}(\mathrm{ZT}_{\mathrm{TEP}})$<br>$= \frac{1}{\Delta \mathrm{T}_{\mathrm{fig}}} \int \mathrm{ZT}_{\mathrm{fig}} \mathrm{dT} - \frac{1}{\Delta \mathrm{T}_{\mathrm{TEP}}} \int \mathrm{ZT}_{\mathrm{TEP}} \mathrm{dT}$ | Bias error<br>Interpolation error | $> 0.1$ |
| Peak ZT filter | $\mathrm{Peak}(\mathrm{ZT}_{\mathrm{fig}}) - \mathrm{Peak}(\mathrm{ZT}_{\mathrm{TEP}})$ | Bias error<br>Extrapolation error | $> 0.1$ |
| Max ZT error filter | $\mathrm{Max}(\delta(\mathrm{ZT}))$ | Bias error<br>Interpolation error<br>Digitisation noise | $> 0.1$ |
| RMS ZT error filter | $\mathrm{Rms}(\delta(\mathrm{ZT})) = \left[\frac{\int \{\delta(ZT)\}^2\ dT}{\int dT}\right]^{\frac{1}{2}}$ | Digitisation noise | $> 0.1$ |
| Normalised Max ZT error with respect to Avg ZT | $\frac{\mathrm{Max}(\delta(\mathrm{ZT}))}{\mathrm{Avg}(\mathrm{ZT}_{\mathrm{TEP}})}$ | Bias error<br>Interpolation error<br>Digitisation noise | $> 0.2$ |
| Normalised Max ZT error with respect to Peak ZT | $\frac{\mathrm{Max}(\delta(\mathrm{ZT}))}{\mathrm{Peak}(\mathrm{ZT}_{\mathrm{TEP}})}$ | Bias error<br>Interpolation error<br>Digitisation noise | $> 0.2$ |

**Table 2. Structure and contents of the teMatDb272 dataset.** Summary of the file types, filenames, field descriptors, and corresponding data content descriptions, including metadata, raw digitised TEP data, and collocated TEP data.

| DB | File type, Filename | Descriptor | Contents |
|---|---|---|---|
| teMatDb272 | File for sample metadata / teMatDb_samples.csv | sample_id | ID of digitised sample |
| | | YEAR | Year of publication |
| | | DOI | DOI of paper |
| | | figure_number_of_targetZT | Figure number showing ZT in the publication |
| | | label_of_targetZT_in_figure | Sample label in the original figure |
| | | figure_label_description | Additional label information |
| | | mat_dimension(bulk, film, 1D, 2D) | Sample's material dimension |
| | | GROUP | Material group labelling |
| | | BASEMAT | Base composition for matrix |
| | | Composition_by_element | Elemental composition |
| | | Composition_detailed | Full stoichiometric composition |
| | | SINTERING | Processing or sintering method |
| | File for raw TEPs from figures / teMatDb_rawTEPs.csv | sample_id | Same as above |
| | | tepname | alpha, rho, kappa for α, ρ, κ<br>ZT for figure of merit |
| | | Temperature | Measured temperature |
| | | tepvalue | TEP value at T |
| | File for collocated TEPs (formatted T, interpolated, no-extrapolated) / teMatDb_collocatedTEPs.csv | sample_id | Same as above |
| | | Temperature | Collocated temperature |
| | | alpha | α |
| | | rho | ρ |
| | | kappa | κ |
| | | ZT_author_declared | ZT value reported in the original figure |
| | File for DB report / z_teMatDb_report | String description | DB info and statistical values |

**Table 3 Summary of teMatDb272 information and key statistical values.** Includes database update time, Sc-ZT filtering criteria, temperature interpolation resolution, and entry counts for TEPs.

| Section | Key | Details or values |
|---|---|---|
| teMatDb272 info | DB updated | 2025-05-15 13:03:44.865946 |
| | formattedDate | 20250515_134730 |
| | dbname | teMatDb |
| | dbversion | v1.1.6 |
| | db_mother | teMatDb272_v1.1.6 |
| DB stats | scZT_filter | criteria_10_10_10_10_20_20 |
| | dT unit for temp collocation | Every 2 K |
| | Number of samples | 272 |
| | Number of publications or DOIs | 262 |
| | Total entries for rawTEPs | 14717 |
| | Entries for $\alpha$ in rawTEPs | 3853 |
| | Entries for $\rho$ in rawTEPs | 3853 |
| | Entries for $\kappa$ in rawTEPs | |
| | Total entries for collocatedTEPs | 56641 |

**Table 4. Number of samples and data quality after Sc-ZT filtering.**

| | Sc-ZT filtering criteria | Number of samples | Data quality in Q-Q plot (high is better) | | |
|---|---|---|---|---|---|
| | | | $R^2$ for $\delta$(ZT) | $R^2$ for $\delta$(Avg ZT) | $R^2$ for $\delta$(Peak ZT) |
| Before filtering | No filter | 355 | 0.6864 | 0.7497 | 0.8296 |
| After filtering | (0.1, 0.1, 0.1, 0.1, 0.2, 0.2)<br>Corresponding to **teMatDb272**, | 272 | 0.9324 | 0.8596 | 0.9510 |
| | (0.05, 0.05, 0.05, 0.05, 0.1, 0.1) | 187 | 0.9799 | 0.9557 | 0.9775 |
| | (0.02, 0.02, 0.02, 0.02 0.04, 0.04) | 71 | 0.9851 | 0.9755 | 0.9845 |

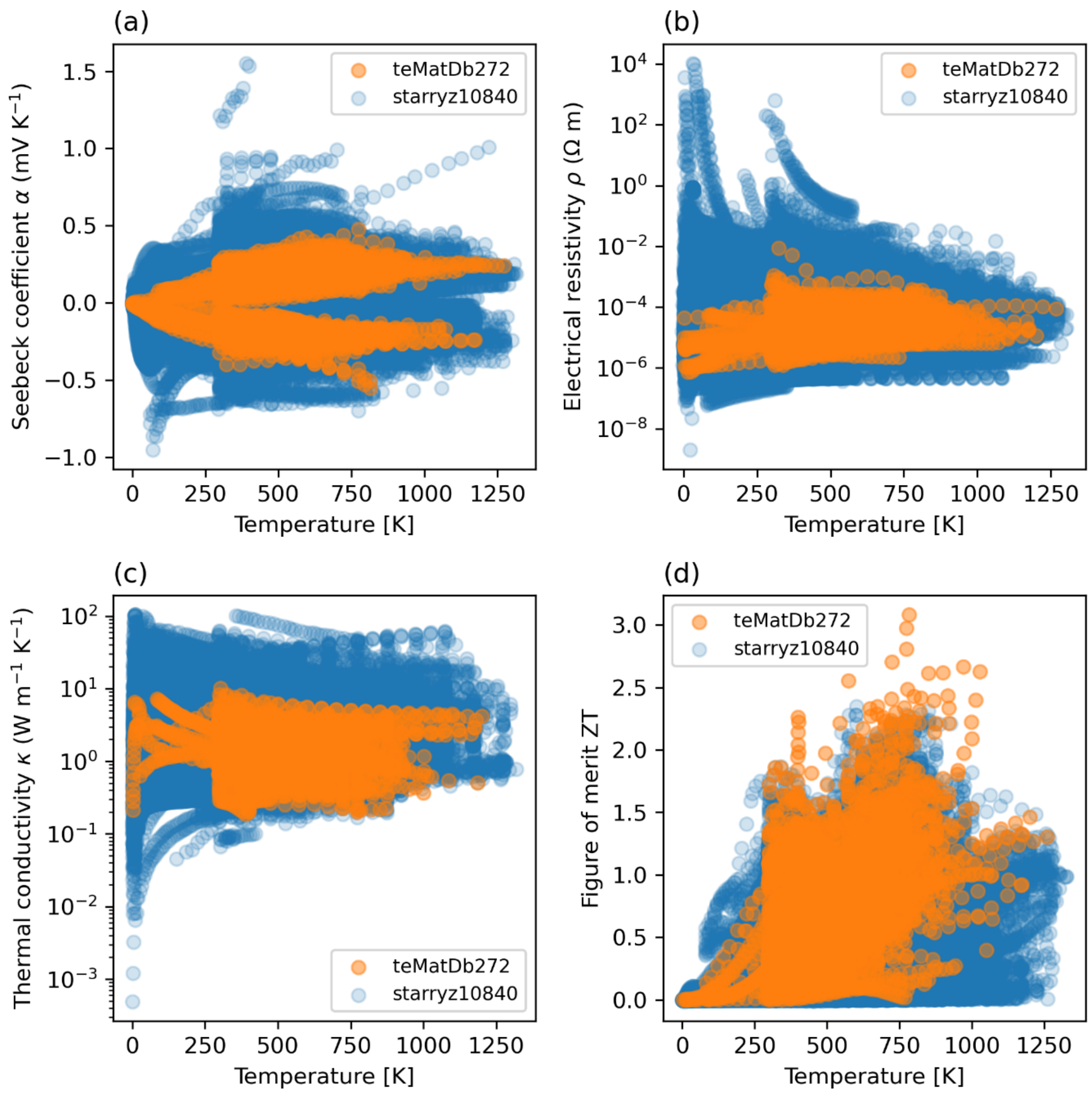

**Figure 1.** **teMatDb272** and thermoelectric property distribution. Curated raw data for (a) the Seebeck coefficient, (b) electrical resistivity, (c) thermal conductivity, and (d) figure of merit are represented and compared to **starryz10840**.

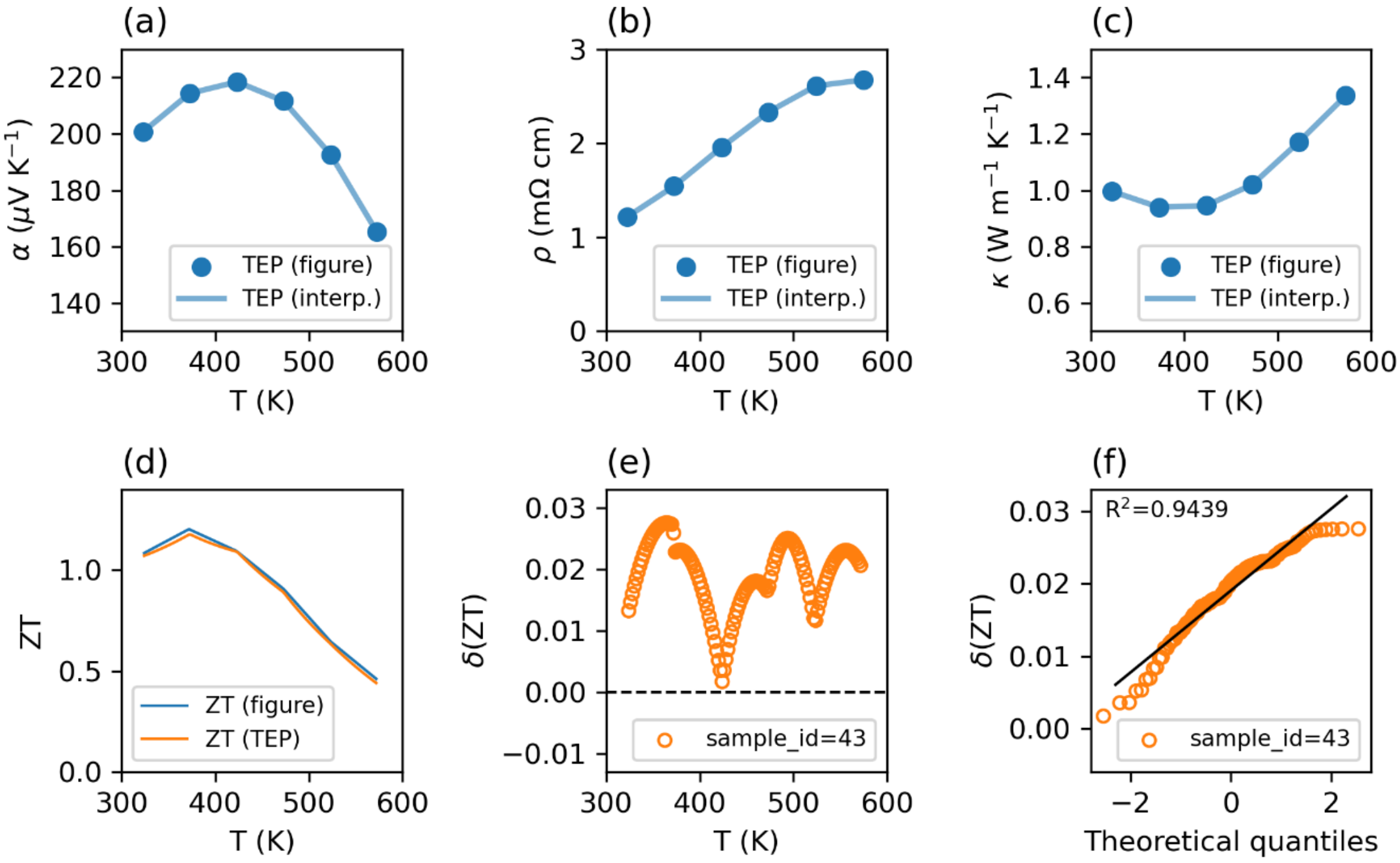

**Figure 2. Digitised thermoelectric property curves for sample_id = 43 in teMatDb272.** (a-c) show the Seebeck coefficient, (b) electrical resistivity, and (c) thermal conductivity, respectively. In (d), the ZT values from the figure are compared with the ZT values recalculated from TEP curves. (e) shows the ZT error as a function of temperature. In (f), the quantile-quantile (Q-Q) plot of $\delta(\mathrm{ZT})$ is shown against the theoretical normal distribution where the y-axis is sorted by the magnitude of the ZT errors. The high $R^2$ value indicates high consistency between the reported ZT from the figure and reported TEP values for this sample.

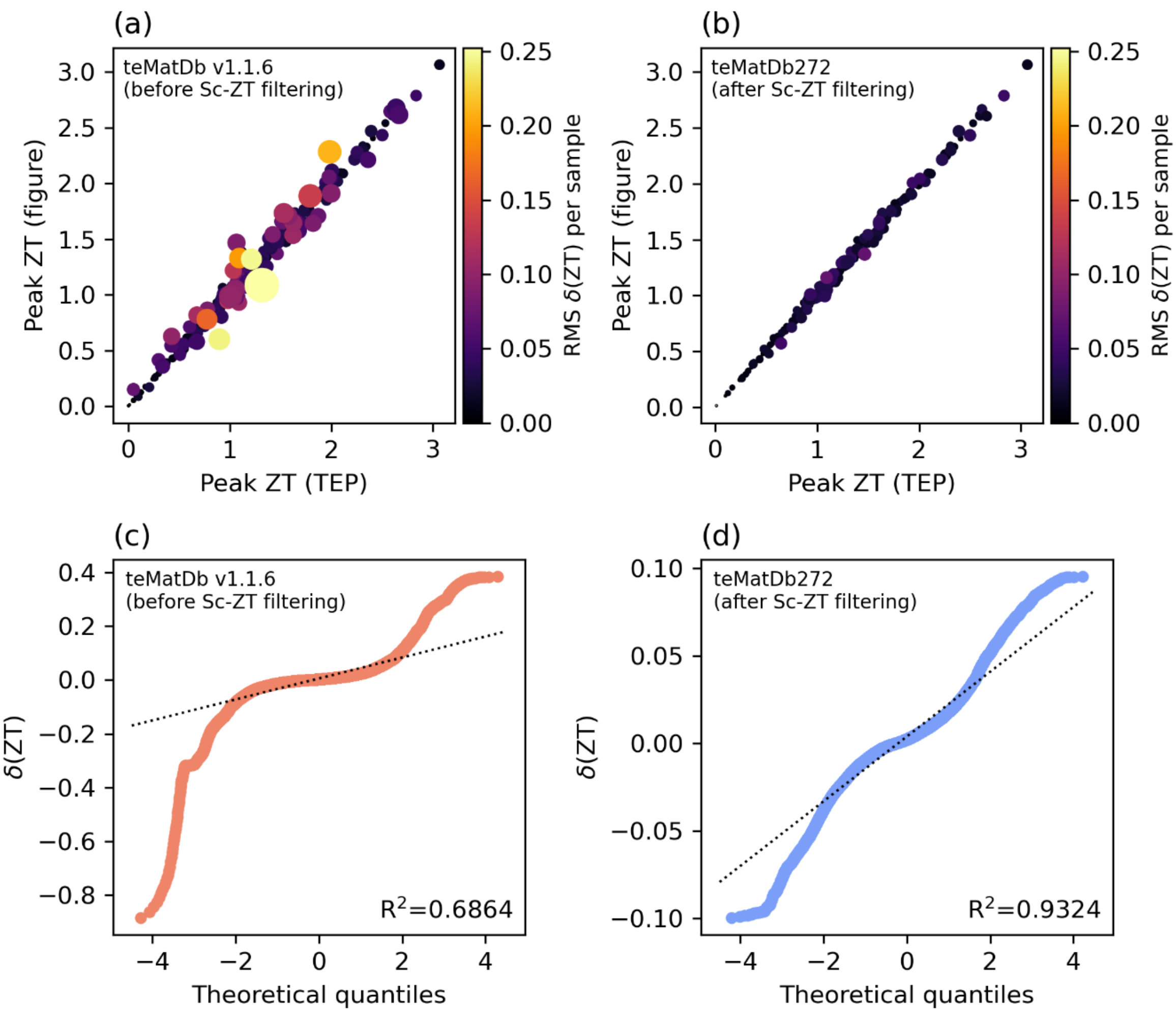

**Figure 3. Evaluation of ZT consistency.** (a) Peak $ZT_{fig}$ values are compared with peak $ZT_{TEP}$ before filtering and (b) after filtering (teMatDb272). (c) Q-Q plot of $\delta(ZT)$ for samples and temperatures before filtering, and (d) Q-Q plot after filtering.

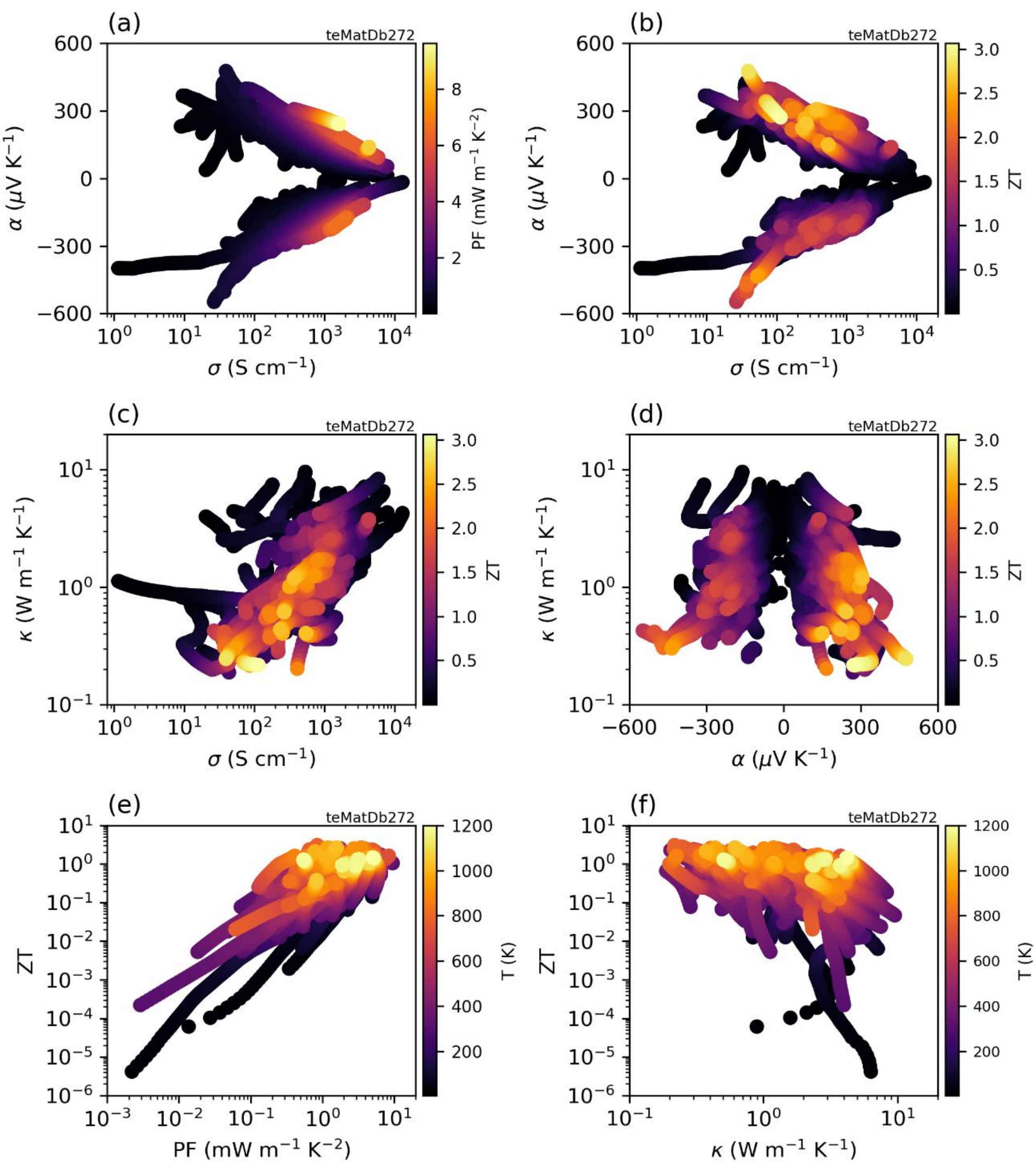

**Figure 4. TEP-TEP distribution.** (a) α-σ scatter plot with PF colour, (b) α-σ with ZT, (c) κ-σ with ZT, (d) κ-α with ZT, (e) ZT-PF with T, and (f) ZT-κ with T.

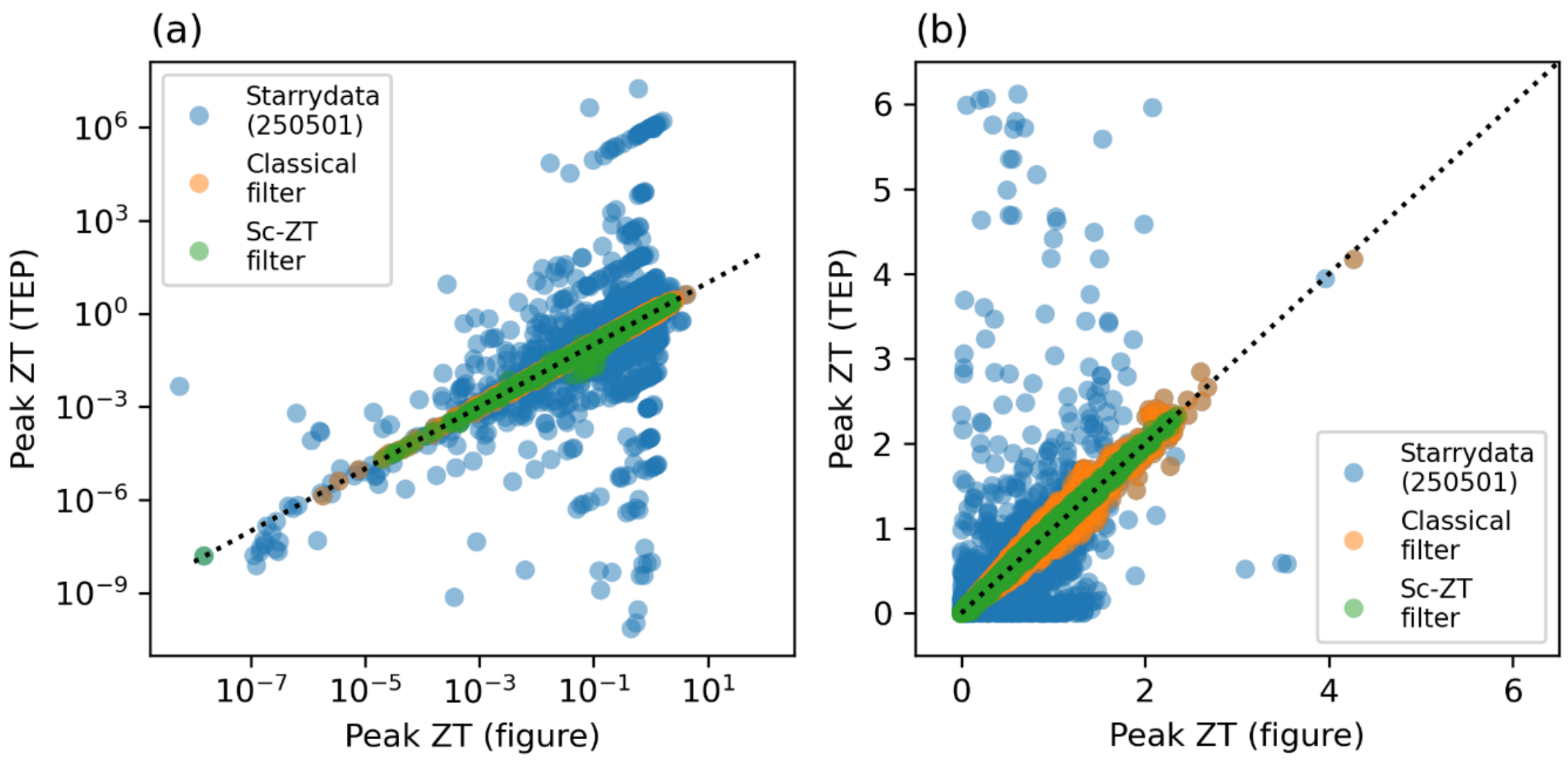

**<span style="color:blue">Figure 5.</span> Peak ZT comparison and filtering effects on Starrydata2 (version: 20250501).** (a) ZT-ZT plots between $ZT_{TEP}$ and $ZT_{fig}$ across Starrydata2 dataset (blue), after Classical filtering (orange), and after Sc-ZT filtering (green). The plot uses a logarithmic scale to highlight large discrepancies caused by unit errors. (b). Zoomed-in linear-scale view of the same data.

# Supporting Information

## Supporting Tables S1-S3

## Supporting Figures S1-S10

**Table S1.** Digitisation error quantification with resolutions.

| | normal | normal | normal | normal |
|---|---|---|---|---|
| | dx_err | dy_err | dx_err | dy_err |
| meanErr | 0.06184 | 0.05136 | 0.03854 | 0.30334 |
| maxErr | 0.14444 | 0.15655 | 0.15734 | 0.41086 |
| err stdev.s | 0.0499 | 0.05194 | 0.05032 | 0.05582 |
| | | | | |
| (digitized and reevaluated values) | | | | |
| **peak value** | **99.4298** | **97.0516** | **99.3585** | **98.4439** |
| rel meanErr | 0.06% | 0.05% | 0.04% | 0.31% |
| rel maxErr | 0.15% | 0.16% | 0.16% | 0.42% |
| rel stdev.s | 0.05% | 0.05% | 0.05% | 0.06% |
| | | | | |
| **avg value** | **51.2022** | **48.2828** | **49.8361** | **50.7999** |
| rel meanErr | 0.12% | 0.11% | 0.08% | 0.60% |
| rel maxErr | 0.28% | 0.32% | 0.32% | 0.81% |
| rel stdev.s | 0.10% | 0.11% | 0.10% | 0.11% |

| | dpi low | dpi low | dpi low | dpi low |
|---|---|---|---|---|
| | dx_err | dy_err | dx_err | dy_err |
| meanErr | 0.08501 | -0.0326 | 0.06559 | 0.41036 |
| maxErr | 0.42392 | 0.52408 | 0.29355 | 0.89299 |
| err stdev.s | 0.20502 | 0.28733 | 0.17646 | 0.26219 |
| | | | | |
| (digitized and reevaluated values) | | | | |
| **peak value** | **99.5416** | **97.0306** | **99.4652** | **98.2533** |
| rel meanErr | 0.09% | -0.03% | 0.07% | 0.42% |
| rel maxErr | 0.43% | 0.54% | 0.30% | 0.91% |
| rel stdev.s | 0.21% | 0.30% | 0.18% | 0.27% |
| | | | | |
| **avg value** | **51.179** | **48.3668** | **49.809** | **50.6929** |
| rel meanErr | 0.17% | -0.07% | 0.13% | 0.81% |
| rel maxErr | 0.83% | 1.08% | 0.59% | 1.76% |
| rel stdev.s | 0.40% | 0.59% | 0.35% | 0.52% |

**Table S2.** Digitisation error quantification with reciprocal parameters.

| | normal | normal | normal | normal |
|---|---|---|---|---|
| | dx_err | dy_err | dx_err | dy_err |
| meanErr | 0.06184 | 0.05136 | 0.03854 | 0.30334 |
| maxErr | 0.14444 | 0.15655 | 0.15734 | 0.41086 |
| err stdev.s | 0.0499 | 0.05194 | 0.05032 | 0.05582 |
| | | | | |
| (digitized and reevaluated values) | | | | |
| **peak value** | **99.4298** | **97.0516** | **99.3585** | **98.4439** |
| rel meanErr | 0.06% | 0.05% | 0.04% | 0.31% |
| rel maxErr | 0.15% | 0.16% | 0.16% | 0.42% |
| rel stdev.s | 0.05% | 0.05% | 0.05% | 0.06% |
| | | | | |
| **avg value** | **51.2022** | **48.2828** | **49.8361** | **50.7999** |
| rel meanErr | 0.12% | 0.11% | 0.08% | 0.60% |
| rel maxErr | 0.28% | 0.32% | 0.32% | 0.81% |
| rel stdev.s | 0.10% | 0.11% | 0.10% | 0.11% |

| | reci | reci | reci | reci |
|---|---|---|---|---|
| | dx_err | dy_err | dx_err | dy_err |
| meanErr | -0.785 | -0.0321 | -0.0182 | -0.0925 |
| maxErr | 13.9 | 0.67013 | 0.40118 | 0.87124 |
| err stdev.s | 2.96217 | 0.12345 | 0.07291 | 0.21241 |
| | | | | |
| (digitized and reevaluated values) | | | | |
| **peak value** | **175.375** | **26.5435** | **20.3333** | **18.2239** |
| rel meanErr | -0.45% | -0.12% | -0.09% | -0.51% |
| rel maxErr | 7.93% | 2.52% | 1.97% | 4.78% |
| rel stdev.s | 1.69% | 0.47% | 0.36% | 1.17% |
| | | | | |
| **avg value** | **12.5443** | **4.65372** | **3.60574** | **3.85644** |
| rel meanErr | -6.26% | -0.69% | -0.51% | -2.40% |
| rel maxErr | 110.81% | 14.40% | 11.13% | 22.59% |
| rel stdev.s | 23.61% | 2.65% | 2.02% | 5.51% |

**Table S3.** Statistics for starryz datasets: starryz10840, starryz15053, and starryz15532.

| db_publication_id | starryz10840 | starryz15053 | starryz15532 |
|---|---|---|---|
| formattedDate | 20250515_234641 | 20250515_234744 | 20250515_234922 |
| dbname | starryz | | |
| dbversion | 20250501_rawdata | | |
| db_mother | starrydata_dataset_250501-0300 | | |
| db_full_id | starryz10840-starrydata_dataset_250501-0300 | starryz15053-starrydata_dataset_250501-0300 | starryz15532-starrydata_dataset_250501-0300 |
| filter | cri_product_def | classic_all_filters | pykeri_TEPZT_readable |
| scZT filter criteria | criteria_10_10_10_10_20_20 | classic_all_filters | pykeri_TEPZT_readable |
| num_sample_id (samples) | 10840 | 15053 | 15532 |
| num_DOI (papers or SID) | 2721 | 3444 | 3293 |
| len_rawTEPs | 710663 | 883707 | 1079894 |
| len_rawalpha | 162934 | 202714 | 249659 |
| len_rawrho | 169982 | 211123 | 260506 |
| len_rawkappa | 141050 | 176681 | 219186 |
| len_rawZT | 142571 | 178127 | 215982 |
| len_collocatedTEPs | 2549971 | 3147078 | 3697834 |
| dT_unit for temp collocation | 4 | | |

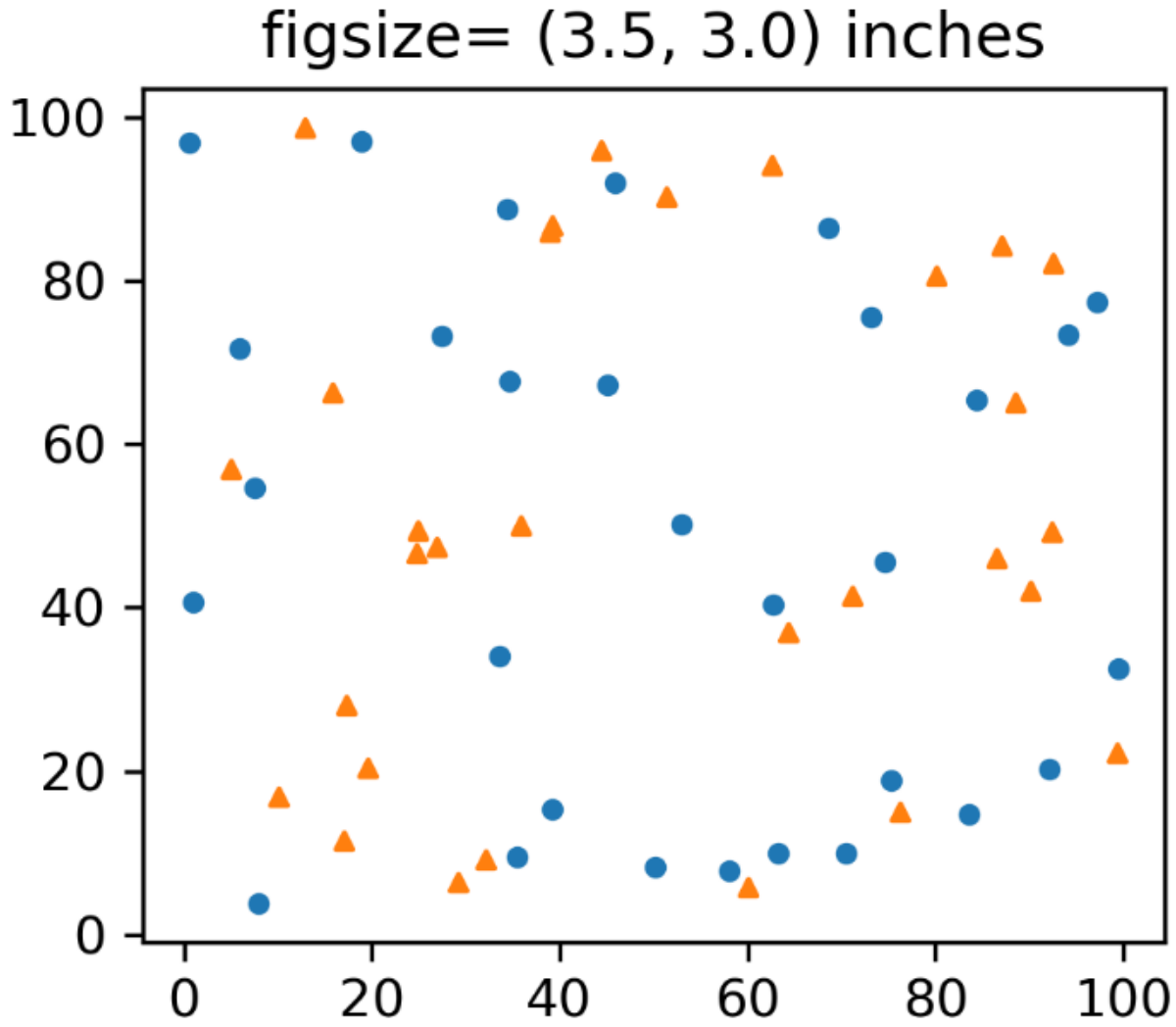

**Figure S1. Generated figure to test digitisation quality.**

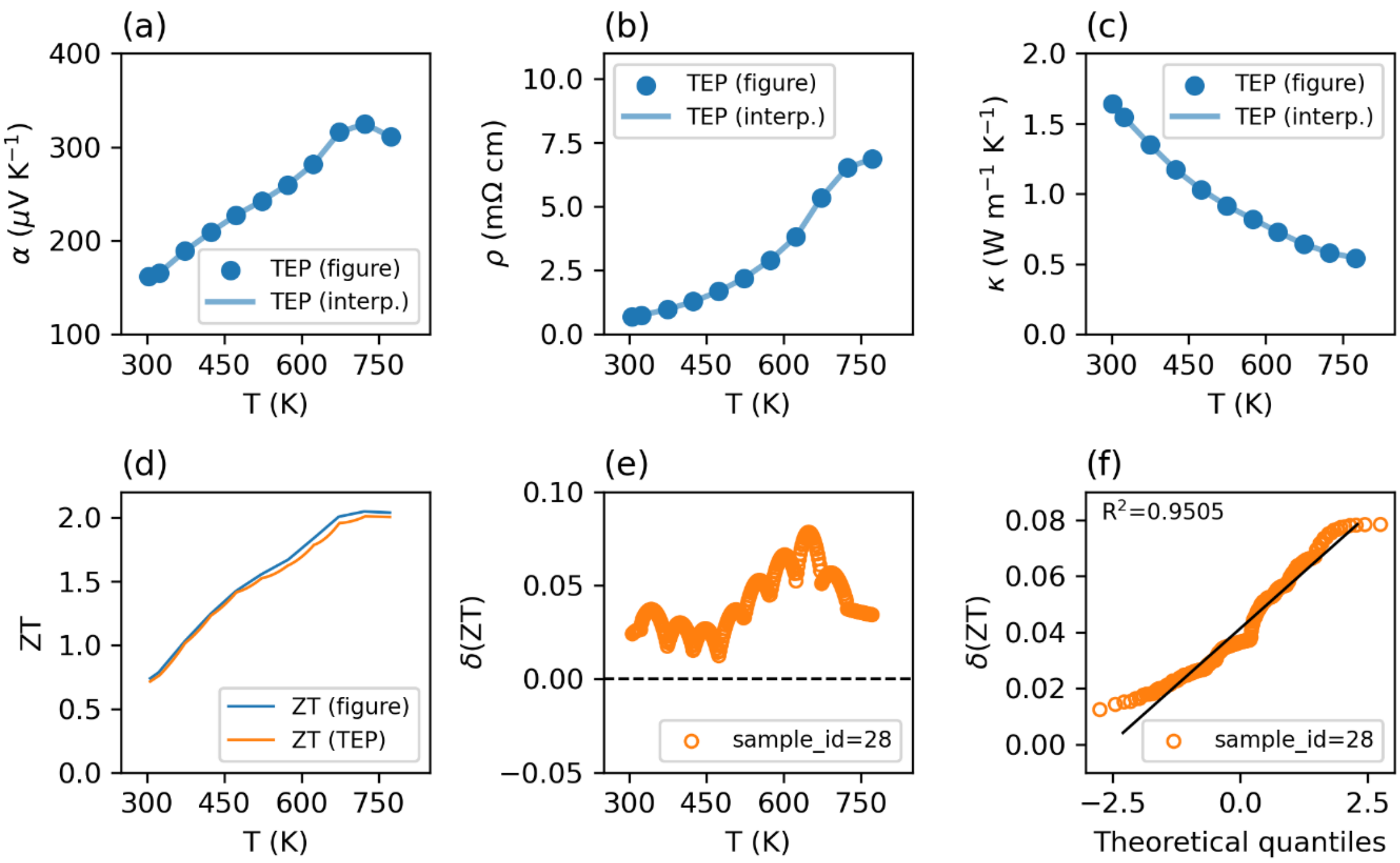

**Figure S2.** **Digitised thermoelectric property curves for sample_id = 28 in teMatDb v1.1.6 and corresponding ZT errors.**

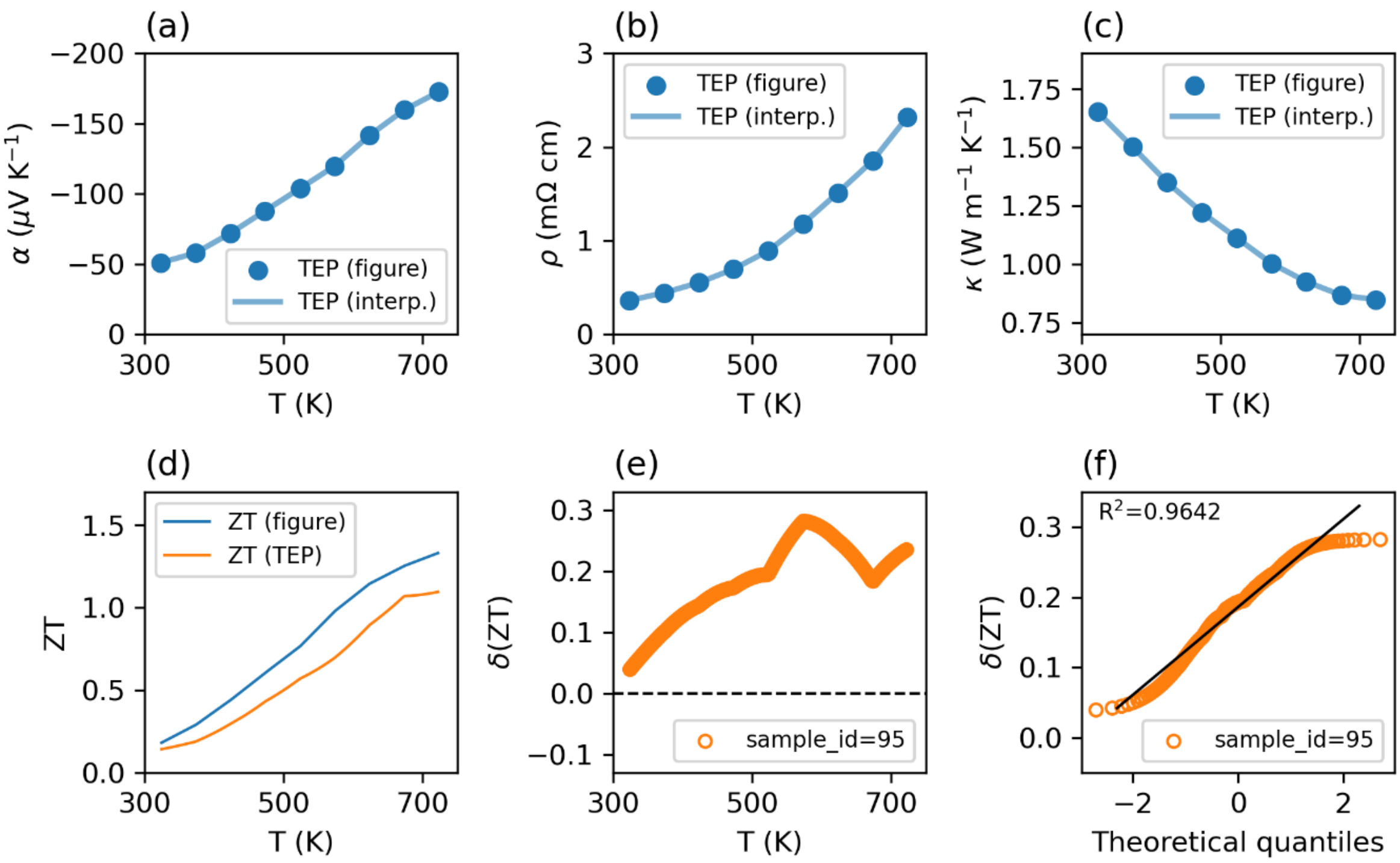

**Figure S3.** **Digitised thermoelectric property curves for sample_id = 95 in teMatDb v1.1.6 and corresponding ZT errors.**

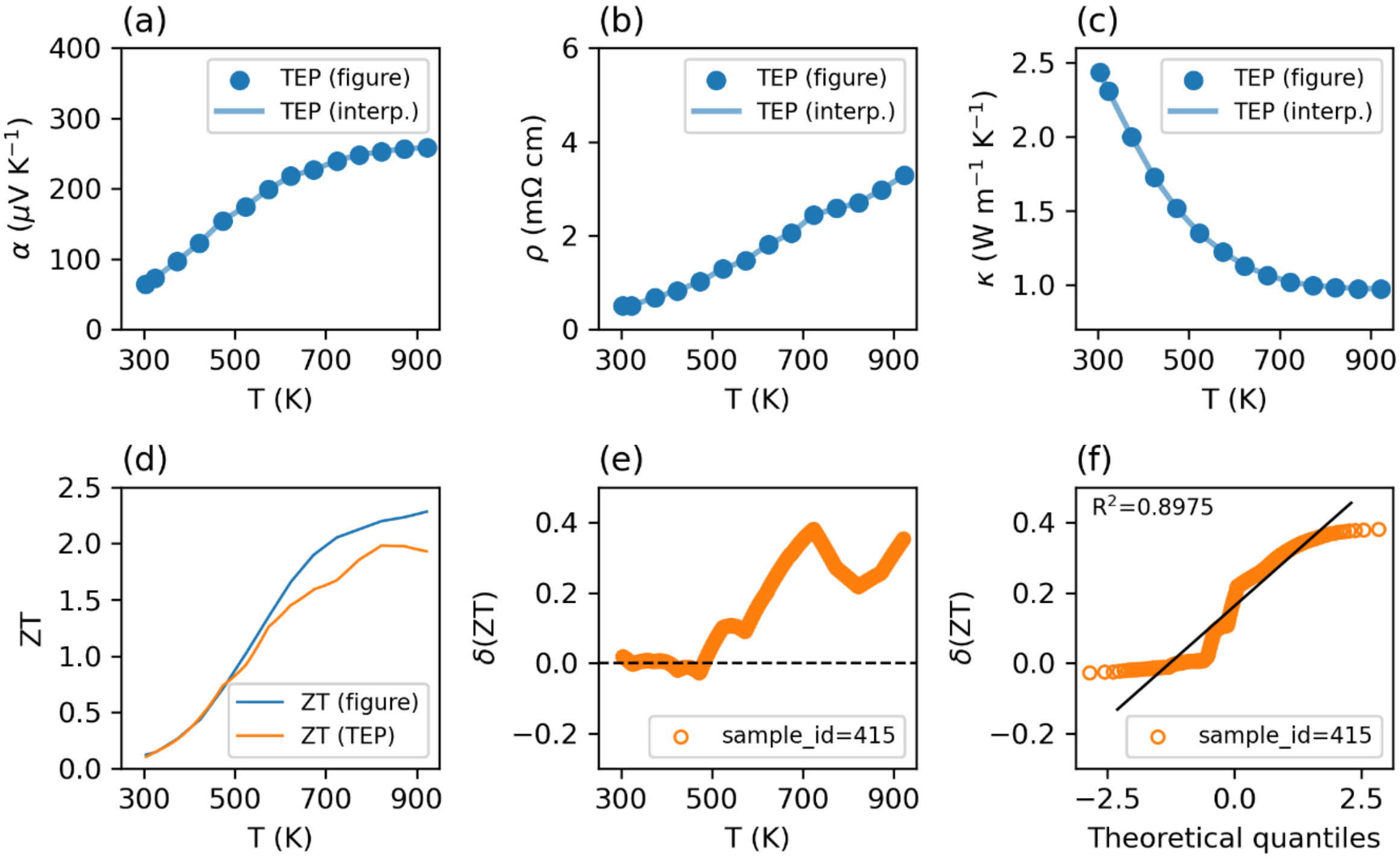

**Figure S4.** **Digitised thermoelectric property curves for sample_id = 415 in teMatDb v1.1.6 and corresponding ZT errors.**

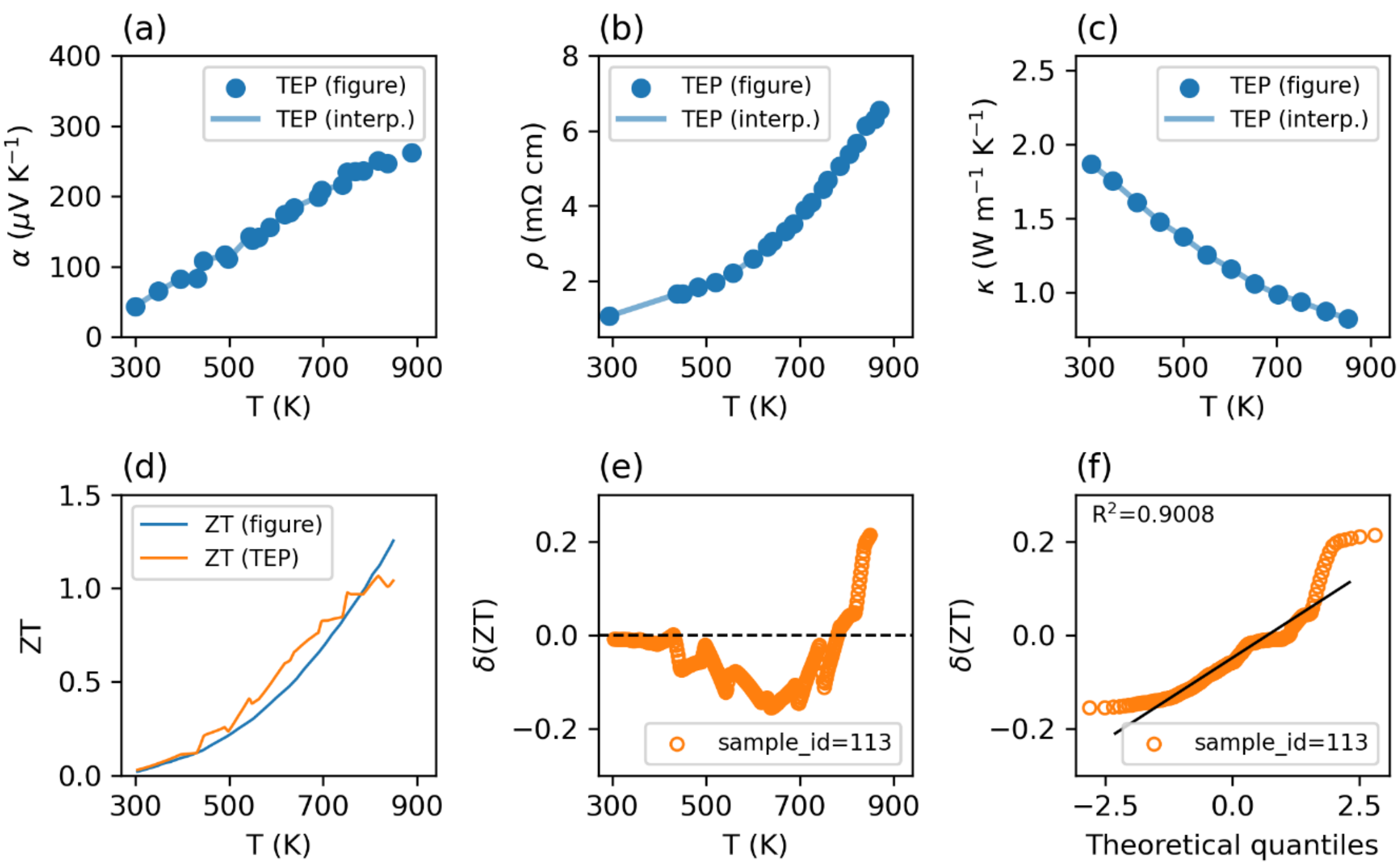

**Figure S5. Digitised thermoelectric property curves for sample_id = 113 in teMatDb v1.1.6 and corresponding ZT errors.**

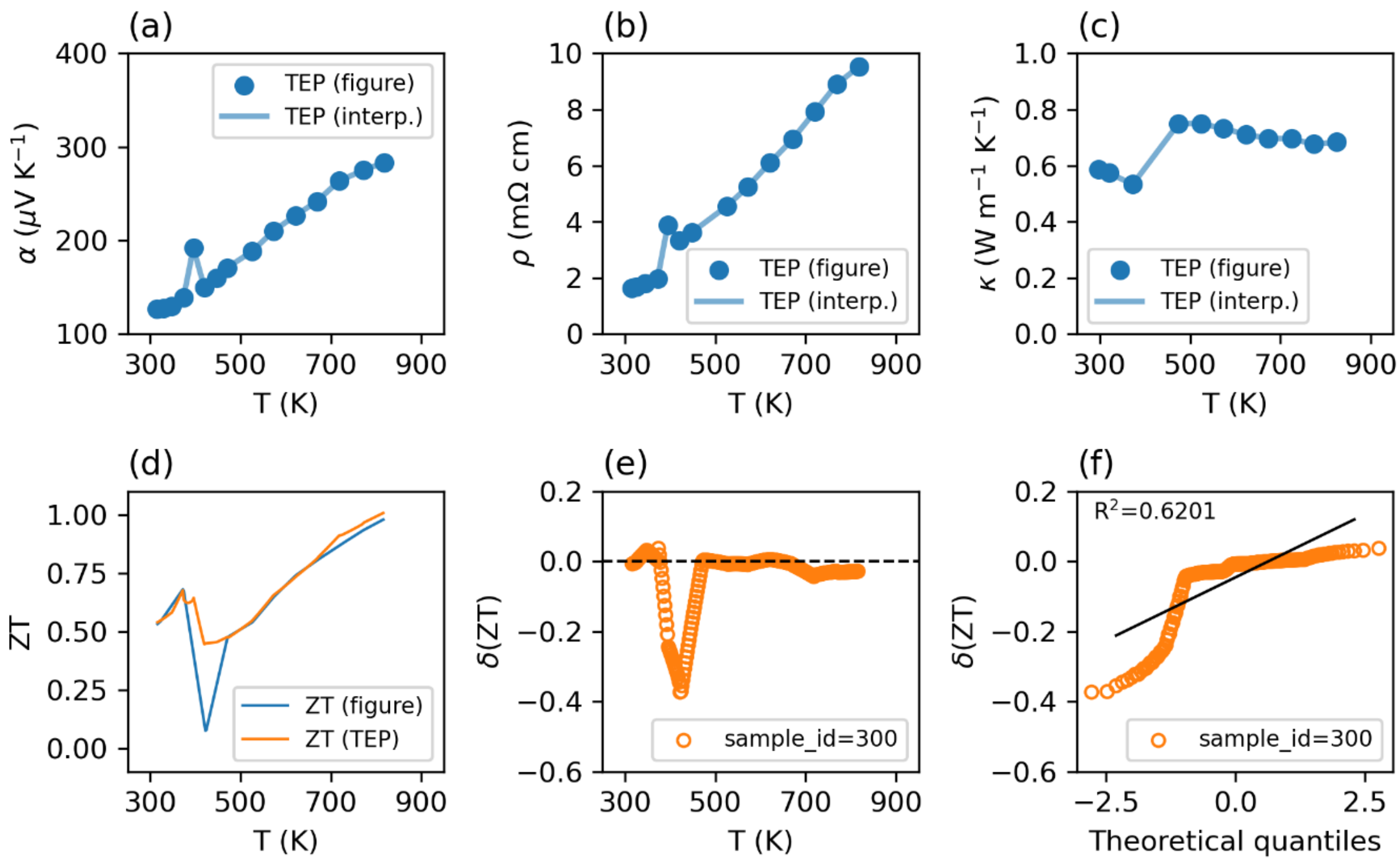

**Figure S6. Digitised thermoelectric property curves for sample_id = 300 in teMatDb v1.1.6 and corresponding ZT errors.**

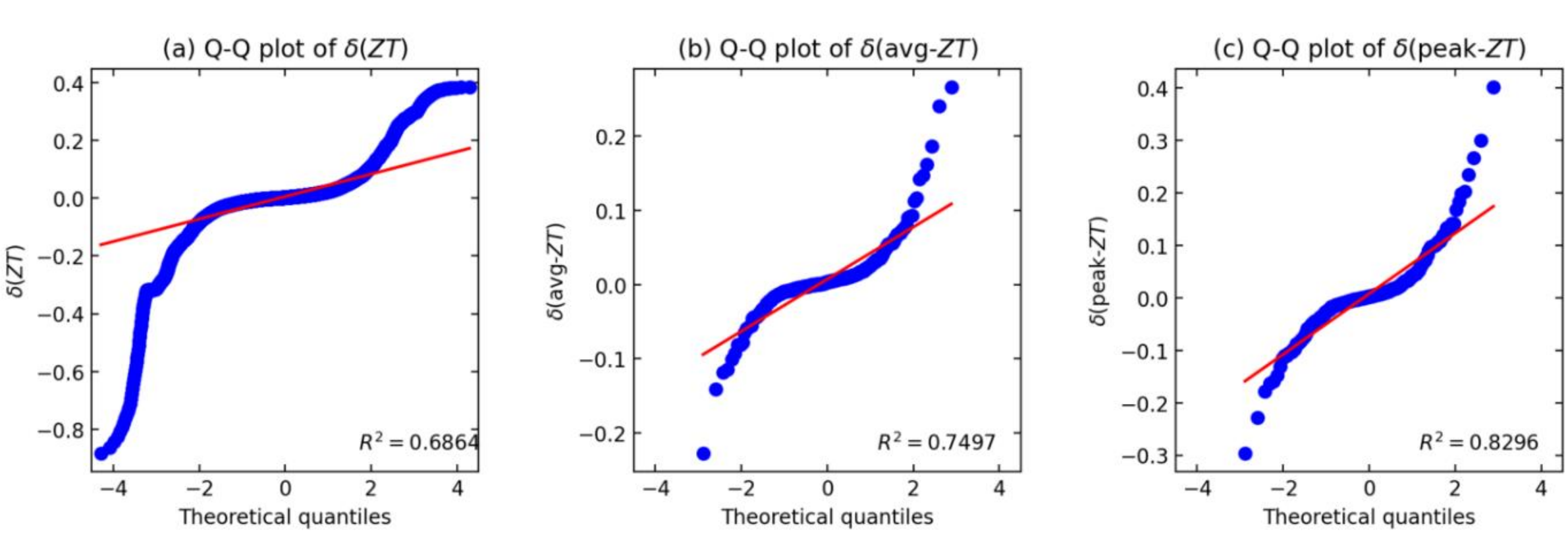

**Figure S7.** **Q-Q plots for teMatDb v1.1.6 before filtering. It has 355 samples.**

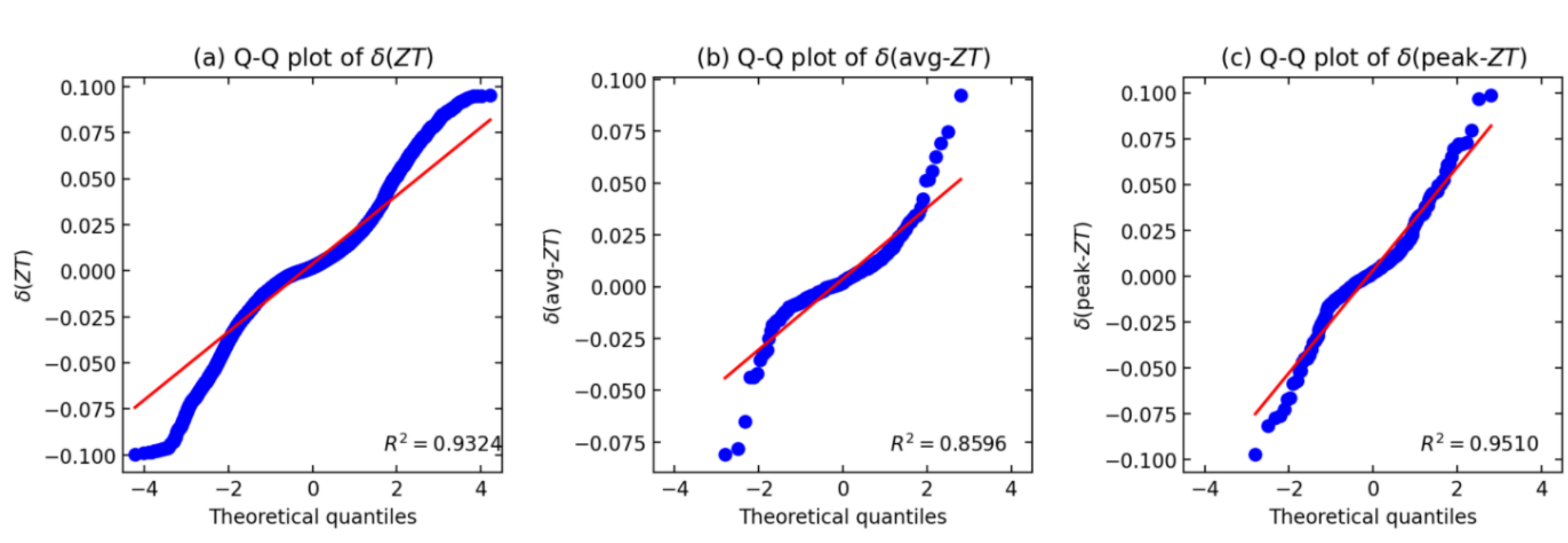

**Figure S8.** **Q-Q plots for teMatDb v1.1.6 after filtering with Sc-ZT criteria of (0.1, 0.1, 0.1, 0.1, 0.2, 0.2), corresponding to teMatDb272. It has 272 filtered samples.**

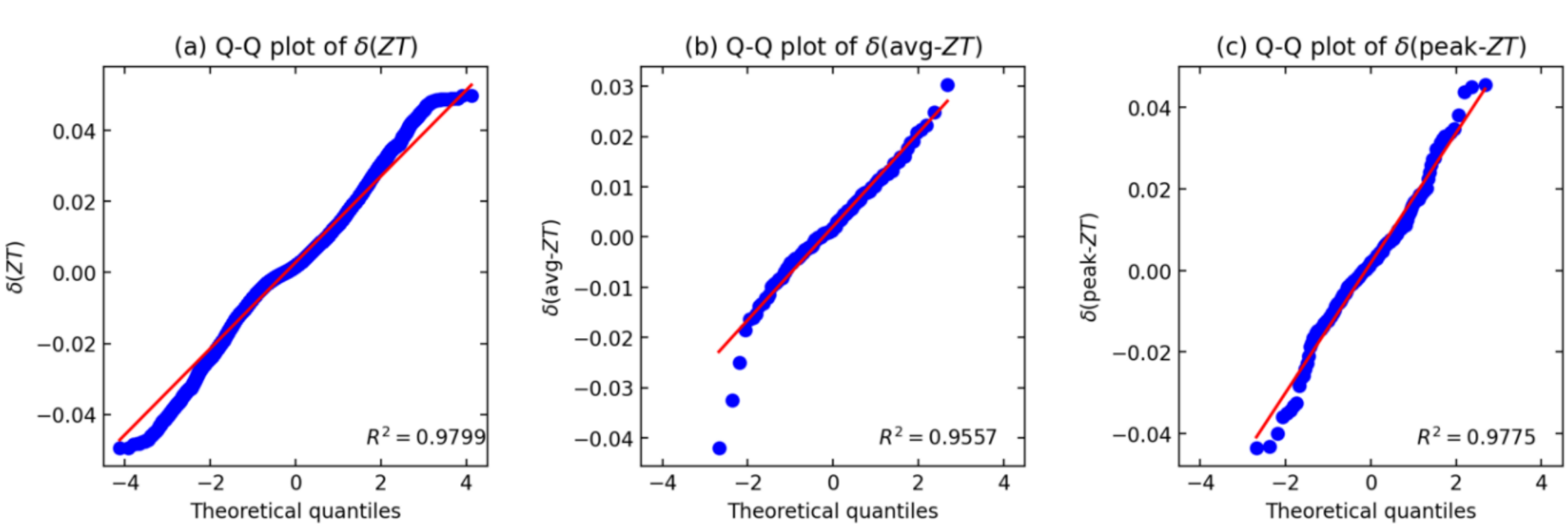

**Figure S9.** **Q-Q plots for teMatDb v1.1.6 after filtering with Sc-ZT criteria of (0.05, 0.05, 0.05, 0.05, 0.1, 0.1). It has 187 filtered samples.**

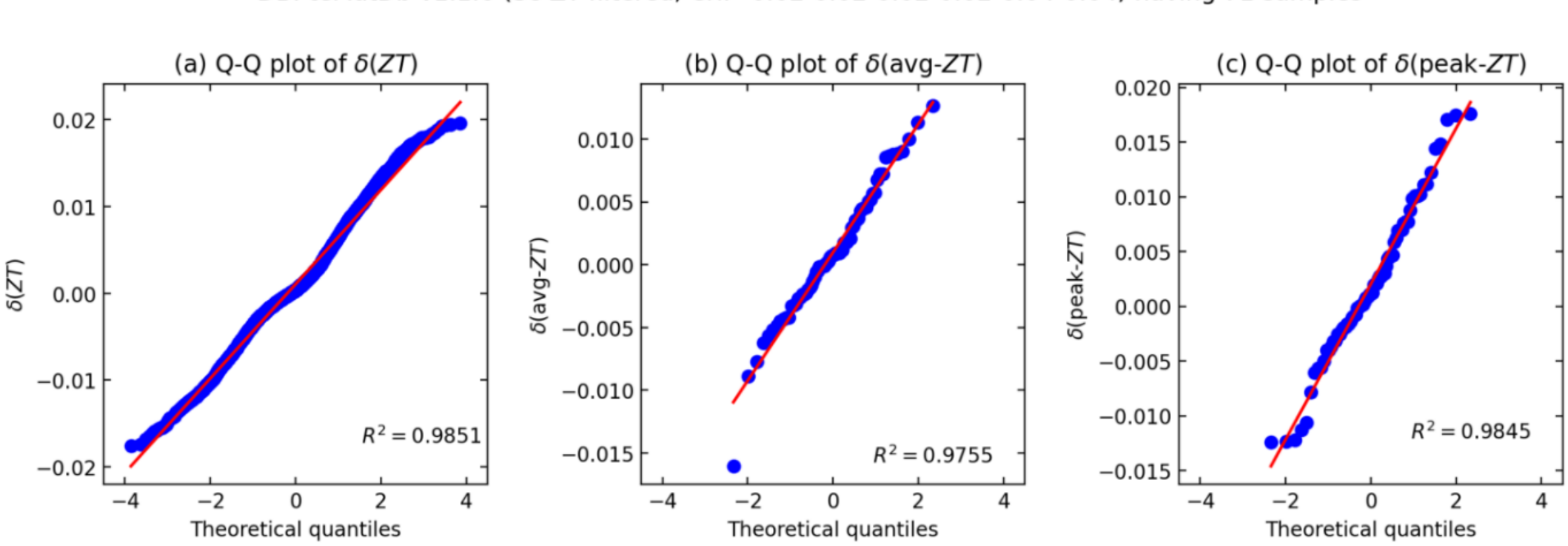

**Figure S10.** **Q-Q plots for teMatDb v1.1.6 after filtering with Sc-ZT criteria of (0.02, 0.02, 0.02, 0.02, 0.04, 0.04). It has 71 filtered samples.**